\documentclass[12pt]{article}
\usepackage{amsmath,amssymb,latexsym,epsf,epsfig,graphicx,tikz}
\pdfoutput=1

\begin{document}

\newcommand{\sect}[1]{\setcounter{equation}{0}\section{#1}}
\renewcommand{\theequation}{\thesection.\arabic{equation}}
\newcommand{\prt}{\partial}
\newcommand{\II}{\mbox{${\mathbb I}$}}
\newcommand{\CC}{\mbox{${\mathbb C}$}}
\newcommand{\RR}{\mbox{${\mathbb R}$}}
\newcommand{\QQ}{\mbox{${\mathbb Q}$}}
\newcommand{\ZZ}{\mbox{${\mathbb Z}$}}
\newcommand{\NN}{\mbox{${\mathbb N}$}}
\def\G{\mathbb G}
\def\UU{\mathbb U}
\def\S{\mathbb S}
\def\tS{\widetilde{\mathbb S}}
\def\A{\mathbb A}
\def\E{\mathbb E}
\def\B{\mathbb B}
\def\D{\mathbb D}
\def\V{\mathbb V}
\def\tV{\widetilde{\mathbb V}}
\newcommand{\cD}{{\cal D}}
\def\hint{H_{\rm int}}

\newcommand{\rd}{{\rm d}}
\newcommand{\diag}{{\rm diag}}
\newcommand{\U}{{\cal U}}
\newcommand{\cP}{{\cal P}}

\newcommand{\ph}{\varphi}
\newcommand{\phd}{\widetilde{\varphi}} 
\newcommand{\phs}{\varphi^{(s)}}
\newcommand{\phb}{\varphi^{(b)}}
\newcommand{\phds}{\widetilde{\varphi}^{(s)}}
\newcommand{\phdb}{\widetilde{\varphi}^{(b)}}
\newcommand{\lambdad}{\widetilde{\lambda}}
\newcommand{\tx}{\widetilde{x}} 
\newcommand{\phl}{\varphi_{i,L}}
\newcommand{\phr}{\varphi_{i,R}}
\newcommand{\phz}{\varphi_{i,Z}}
\newcommand{\mur}{\mu_{{}_R}}
\newcommand{\mul}{\mu_{{}_L}}
\newcommand{\muv}{\mu_{{}_V}}
\newcommand{\mua}{\mu_{{}_A}}

\def\a{\alpha}

\def\cA{\mathcal A} 
\def\C{\mathcal C} 
\def\T{\mathcal T}
\def\O{\mathcal O}
\def\N{\mathcal N}
\def\I{\mathcal I}
\def\der{\partial }
\def\mis{{\frac{\rd k}{2\pi} }}
\def\ri{{\rm i}}
\def\xt{{\widetilde x}}
\def\ft{{\widetilde f}}
\def\gt{{\widetilde g}}
\def\qt{{\widetilde q}}
\def\tt{{\widetilde t}}
\def\tmu{{\widetilde \mu}}
\def\prt{{\partial}}
\def\tr{{\rm Tr}}
\def\inc{{\rm in}}
\def\out{{\rm out}}
\def\e{{\rm e}}
\def\eps{\varepsilon}
\def\k{\kappa}
\def\v{{\bf v}}
\def\ebf{{\bf e}}
\def\abf{{\bf A}}

\def\be{\begin{equation}}
\def\ee{\end{equation}}

\def\bea{\begin{eqnarray}}
\def\eea{\end{eqnarray}}
\def\a{\alpha}

%%%%%%%%%%%%%%%%%%% INIZIO %%%%%%%%%%%%%%%%%%%%%%%

%%%%%%%% 
\newcommand{\finprf}{\null \hfill {\rule{5pt}{5pt}}\\[2.1ex]\indent}

%%%%%%%%%%%%%%%%%%%%%%%
\pagestyle{empty}
\rightline{January 2012}

\vfill

\begin{center}
{\Large\bf Entanglement Entropy of Quantum Wire Junctions}
\\[2.1em]

\bigskip

{\large
Pasquale Calabrese$^{a,b}$, 
Mihail Mintchev$^{b,a}$ and Ettore Vicari$^{a,b}$}\\

\null

\noindent 

{\it  
$^a$ Dipartimento di Fisica dell'Universit\`a di
Pisa, Largo Pontecorvo 3, 56127 Pisa, Italy\\[2.1ex]
$^b$ Istituto Nazionale di Fisica Nucleare, Largo Pontecorvo 3, 56127 Pisa, Italy}
\vfill

\end{center}
\begin{abstract} 

We consider a fermion gas on a star graph modeling a quantum wire junction and derive the entanglement 
entropy of one edge with respect to the rest of the junction. 
The gas is free in the bulk of the graph, the interaction being localized in its vertex and 
described by a non-trivial scattering matrix. We discuss 
all point-like interactions, which lead to unitary time evolution of the system. We show that for a finite number 
of particles $N$, the R\'enyi entanglement entropies of one edge grow as $\ln N$ with a calculable prefactor, which 
depends not only on the central charge, but also on the total transmission probability from the considered edge to the rest 
of the graph. This result is extended to the case with an harmonic potential in the bulk.

\end{abstract}
\bigskip 
\medskip 
\bigskip 

\vfill
\rightline{IFUP-TH 21/2011}
\newpage
\pagestyle{plain}
\setcounter{page}{1}

%%%%%%%%%%%%%%%%%%%%%%%%%%%%%%%%

\sect{Introduction} 
\bigskip 

Quantum field theory on graphs attracted recently much attention mainly in relation with the study 
\cite{kf-92}-\cite{y-02} of the transport properties of quantum wire networks. Different frameworks 
\cite{lrs-02}-\cite{Soori:2010ga} have been developed to investigate the phase diagram and the conductance of these structures. 
Despite of the fact that the universal properties in the bulk are described by the well known Luttinger liquid theory, 
the different boundary conditions at the junctions lead to exotic phase diagrams \cite{coa-03, emabms-05, ff-05,
Bellazzini:2006kh, hc-08, dr-08, Bellazzini:2008fu, Bellazzini:2009nk} whose degree of 
universality is not completely understood and is still under investigation. The results, concerning the charge 
transport, confirm that the conductance properties of the quantum wire networks are strongly affected by 
the boundary conditions as well. 

In the present paper we analyze another physical quantity - the entanglement entropy of one edge of the junction with 
respect to all the others edges. 
Lots of studies on the entanglement properties of many-body systems in the last decade have unveiled new (universal) features
of these systems and somehow put their global understanding on a deeper level (see e.g. the reviews \cite{rev}).     
In particular,  von Neumann and R\'enyi entanglement entropies of the reduced density
matrix $\rho_A$ of a subsystem $A$ turned out to be particularly useful for 1D systems.  
R\'enyi entanglement entropies are defined as 
\be
S^{(\a)}=\frac1{1-\a}\ln{\rm Tr}\,\rho_{A}^\a\, .
\label{Sndef}
\ee 
For $\a\to1$ this definition gives the most commonly used von
Neumann entropy $S^{(1)}=-\tr{}{\rho_A\ln\rho_A}$, while for $\a\to\infty$
is the logarithm of the largest eigenvalue of $\rho_A$ also known as
single copy entanglement \cite{sce}.
Furthermore, the knowledge of the $S^{(\a)}$ for different $\alpha$ characterizes the
full spectrum of non-zero eigenvalues of $\rho_A$ \cite{cl-08}.

One of the most remarkable results is the universal behavior displayed
by the entanglement entropy at 1D conformal quantum critical points, 
determined by the central charge \cite{c-lec} of the underlying conformal field theory
(CFT) \cite{holzhey,vidalent,cc-04,cc-rev}.  For a partition of an
infinite 1D system into a finite piece $A$ of length $\ell$ and the
remainder, the R\'enyi entanglement entropies for $\ell$ much larger
than the short-distance cutoff $a$ are \be
S^{(\a)} =\frac{c}6\left(1+\frac1\a\right) \ln \frac{\ell}a +c_\a\,,
\label{criticalent}
\ee 
where $c$ is the central charge and $c_\a$ a non-universal constant.

Given the  importance of this result (and also many others not mentioned here) for homogeneous systems, it is natural to wonder whether 
in the case of junctions the entanglement entropies can share some light on the universality and on the relevance of the 
parameters defining the junction. 
Previous studies in the subject \cite{l-04,p-def,def2,isl-09,ss-08,scla-07,ep-10,p-11,eg-10} 
have been limited to the case of only two edges (i.e. an infinite line with a defect) 
and most often performed for lattice models. 
These results provide a strong evidence that the logarithmic behavior in Eq. (\ref{criticalent}) remains valid even in the 
presence of defects,
but the prefactor does not depend only on the central charge of the bulk CFT when the defect is a {\it marginal} perturbation (in 
renormalization group sense) as it is known to happen for free fermions \cite{kf-92}.

In order to tackle the problem of entanglement in a junction with an arbitrary number of wires,  
we use the recently developed systematic framework  \cite{Calabrese:2011zz, Calabrese:2011vh} for calculating the 
bipartite entanglement entropy of spatial subsystems of one-dimensional quantum systems in continuous space. 
We only consider a free fermion gas in bulk in which the junction introduces a marginal perturbation. 
The junction boundary conditions  define a specific scattering matrix $\S$, encoding all possible point-like interactions in
the vertex which give rise to  unitary time evolution. 
Focussing on the scale invariant case, we show that for a finite number 
of particles $N$ and for edges of equal length $L$, the R\'enyi entanglement entropies of any of the edges grow as $\ln N$.
Oppositely to the case in the absence of the point-like interaction (i.e. Eq. (\ref{criticalent})), the
prefactor of this logarithm does not depend only on the central charge, 
but also on the total transmission probability $(1-|\S_{ii}|^2)$ from the considered edge $i$ to the rest of the graph. 
Some of the results presented here have been anticipated in the short communication \cite{Calabrese:2011zz}.
We show also that the presence of an external harmonic potential in the bulk (acting identically on all edges) does not alter this result.

The paper is organized as follows. In the next section we describe the basic features of the model 
and the scattering matrices generated by the point-like interactions at the junction. We discuss in detail 
the scale invariant case and derive the two-point correlation function. 
The entanglement entropy, associated with this system, is analytically computed in section 3. In section 4 we 
extend our considerations, adding a harmonic potential in the bulk. Section 5 is 
devoted to the conclusions and the discussion of some further developments in the subject.

\sect{Schr\"odinger junction}
\bigskip 

\subsection{The general setting} 

\begin{figure}[t]
\begin{center}
\begin{picture}(550,150)(10,905) 
\includegraphics[scale=1.3]{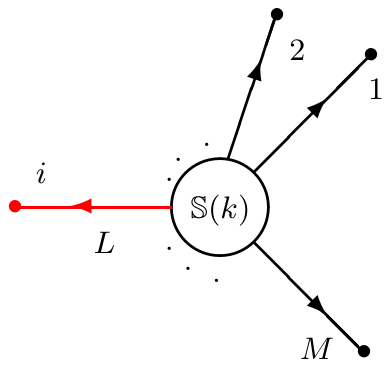}
\end{picture} 
\end{center}
\caption{(Color online) A star graph $\Gamma$ with scattering matrix $\S(k)$ in the vertex and all edges of length $L$. 
We consider the entanglement entropy of the edge $i$ (red) with respect to all the others.} 
\label{Figjunction}
\end{figure}

In this section we consider a gas of $N$ spinless fermions on a quantum wire junction. 
We consider only the ground-state of such system %(i.e. the Fock vacuum) 
and we 
refer to it as the ground-state of $N$ {\it particles}, having in mind that the particles are the 
original fermions and not the excitations above the ground-state (that are usually referred as 
particles in field theory literature). 
A simple model, describing the junction, is represented by a star graph $\Gamma$ with
$M$ edges of finite length $L$, as shown in Fig. \ref{Figjunction}. Each point $P$ in the bulk of
$\Gamma$ is parametrized by $(x,i)$, where $0\leq x\leq L$ is the
distance of $P$ from the vertex $V$ of the graph and $i$ labels the
edge. We assume that in the bulk ($x\not=0$ and $x\not=L$) the gas is free and is described 
by the Schr\"odinger field $\psi_i(t,x)$, which satisfies
\begin{equation}
\left (i \partial_t + \frac{1}{2m} \partial_x^2\right )\psi_i (t,x) = 0 
\label{eqm1}
\end{equation}
and standard equal-time canonical anticommutation relations.  
The only non-trivial interactions are localized in the vertex $V$ of $\Gamma$ and are encoded in 
boundary conditions at $x=0$. These conditions are fixed in turn by imposing that the 
bulk Hamiltonian $-\partial_x^2$ admits a self-adjoint extension on the whole graph. 
In such a way all point-like interactions, leading to a unitary time evolution of the system, are covered. 
The most general boundary conditions, implementing this natural physical requirement, are \cite{ks-00, h-00} 
at the vertex
\begin{equation} 
\sum_{j=1}^{M} \left [\lambda ({\mathbb I}-{\mathbb U})_{ij}\, 
\psi_j (t,0) -i ({\mathbb I}+{\mathbb U})_{ij}
(\partial_x\psi_j ) (t,0)\right ] = 0\, , 
\label{bc1} 
\end{equation}
where ${\mathbb U}$ is an arbitrary $M\times M$ unitary matrix and $\lambda$ a real parameter with the dimension of mass.
To fully specify the problem we also need to impose boundary conditions at the external ends of the edges. 
The most general ones are 
\begin{equation} 
(\partial_x\psi_i ) (t,L) = \mu_i\, \psi_i(t,L) \, , 
\label{bc2} 
\end{equation}
where $\mu_i$ are again real parameters with the dimension of mass. 
Eq. (\ref{bc2}) is the familiar Robin (mixed) boundary condition.  
Notice that Eq. (\ref{bc1}) extends this condition to the vertex $V$ of the graph.

It has been established in \cite{ks-00, h-00} that the point-like interaction, induced by  
(\ref{bc1}), generates the scattering matrix 
\begin{equation} 
{\mathbb S}(k) = -{[\lambda ({\mathbb I} - {\mathbb U}) - 
k({\mathbb I}+{\mathbb U} )]\over [\lambda ({\mathbb I} - {\mathbb U}) 
+ k({\mathbb I}+{\mathbb U} )]}\, . 
\label{S1}
\end{equation} 
Besides of unitarity 
\begin{equation} 
{\mathbb S}(k) {\mathbb S}^\dagger(k) = \II \, , 
\label{S2} 
\end{equation} 
$\S(k)$ satisfies Hermitian analyticity 
\begin{equation} 
{\mathbb S}^\dagger(k) = {\mathbb S}(-k) 
\label{S3} 
\end{equation} 
as well. Notice also that 
\begin{equation} 
{\mathbb S}(\lambda ) = {\mathbb U}\, , 
\quad {\mathbb S}(-\lambda ) = {\mathbb U}^{-1} \, , 
%\label{S2} 
\end{equation} 
showing that the unitary matrix ${\mathbb U}$, entering the boundary conditions
(\ref{bc1}), is actually the scattering matrix at the scale $\lambda$. 

The main difficulty in solving the Schr\"odinger equation (\ref{eqm1}) on the graph $\Gamma$ is the mixing between the different edges 
codified in the boundary conditions (\ref{bc1}) and (\ref{bc2}). 
In order to simplify this problem, we impose that the boundary conditions at the ends of each arm are all the same, in such 
a way to restore (at the level of the Hamiltonian)
permutation symmetry of the edges of the graph. 
It should be clear from the physical point of view that, being interested in the thermodynamic limit with $L,N\to\infty$,
the boundary conditions at $L$ must not affect the final result. 
Thus we assume from now on that 
\begin{equation}
\mu_1=\mu_2= \cdots = \mu_M \equiv \mu \, .  
\label{h1}
\end{equation} 
Under this condition, Eqs.  (\ref{bc1}) and  (\ref{bc2}) can be rewritten in equivalent forms without mixing. Indeed, let us introduce 
the unitary matrix ${\cal U}$ diagonalizing ${\mathbb U}$, namely 
\begin{equation} 
{\cal U}\, {\mathbb U}\, {\cal U}^\dagger = {\mathbb U}_d=  
{\rm diag} \left (e^{-2i\alpha_1}, e^{-2i\alpha_2}, ... , 
e^{-2i\alpha_{M}}\right )\, , 
\qquad -{\pi\over 2} < \alpha_i \leq {\pi\over 2} \, . 
\label{d11}
\end{equation} 
Remarkably enough, ${\cal U}$ diagonalizes also ${\mathbb S}(k)$ for {\it any} $k$:
\begin{equation} 
{\mathbb S}_d(k) = {\cal U}^\dagger {\mathbb S}(k) {\cal U} = \\
{\rm diag} \left ({k+i \eta_1\over 
k-i \eta_1}, {k+i \eta_2\over k-i \eta_2}, ... , 
{k+i \eta_M\over k-i \eta_M} \right ) \, , 
\label{d3}
\end{equation} 
where 
\begin{equation} 
\eta_i \equiv \lambda {\rm tan} (\alpha_i)\, .  
\label{d4}
\end{equation} 
It is quite natural at this point to introduce the fields  
\begin{equation} 
\varphi_i(t,x) = \sum_{j=1}^{M} {\cal U}_{ij} \psi_j(t,x) \, , 
\label{nl1}
\end{equation} 
which obviously satisfy Eq. (\ref{eqm1}).  In terms of $\varphi_i$ the
boundary conditions (\ref{bc1},\ref{bc2}) decouple,
\begin{eqnarray} 
(\partial_x\varphi_i ) (t,0) &=& \eta_i\, \varphi_i(t,0) \, , 
\label{bc3} \\
(\partial_x\varphi_i ) (t,L) &=& \mu \, \varphi_i(t,L) \, , 
\label{bc4} 
\end{eqnarray} 
defining a simple spectral problem on the tensor product 
${\cal H} = \bigotimes_{i=1}^{M} {\mathbb L}^2[0,L]$, which is analyzed below.  

It is worth stressing that $\varphi_i(t,x)$ is a superposition of the values of the original field 
$\psi_i(t,x)$ at the same distance $x$ from the vertex, but on {\it different} edges of the junction. Being so 
delocalized, $\varphi_i(t,x)$ is unphysical and provides only a convenient basis for dealing with the boundary 
conditions. The physical observables and correlation functions will be always expressed in terms of 
the physical fields $\psi_i(t,x)$. 

The eigenfunctions of $-\partial_x^2$, obeying (\ref{bc3}) and (\ref{bc4}) are 
\begin{equation} 
\phi_i(k,x) = c_i \left (e^{i k x} + 
{k+i \eta_i\over k-i \eta_i} e^{-i k x}\right ) \, , 
\qquad k \geq 0\, , 
\label{nl2}
\end{equation} 
where $c_i$ are some constants to be fixed below and $k$ satisfy
\begin{equation} 
e^{2ikL} = \left ({k+i\eta_i\over k-i\eta_i}\right )\left 
( {k-i\mu\over k+i\mu}\right ) \, .  
\label{nl3}
\end{equation} 
In order to determine the spectrum of $k$ explicitly, we simplify the problem further by requiring scale invariance. 
\bigskip 

\subsection{The scale invariant case} 
\medskip 

Scale invariance of the boundary conditions (\ref{bc3}) and (\ref{bc4}) implies the values 
\begin{equation}
\mu = 
\begin{cases} 
0 \, , \\
\infty \, , \\ 
\end{cases} 
\qquad 
\eta_i = 
\begin{cases} 
0 \quad \; \;  (\alpha_i=0)\, , & \qquad  \text{Neumann b.c.}\, , \\
\infty \quad (\alpha_i=\pi /2)\, , & \qquad  \text{Dirichlet b.c.} \\ 
\end{cases} 
\label{si1}
\end{equation} 
In other words, the critical points are fixed by the $(M+1)$ vector
$(\mu, \eta_1,\eta_2,...,\eta_{M})$ whose components take the values (\ref{si1}). For an edge $i$ 
one has the following possibilities: 
\medskip 

(a) $\mu =0$ (Neumann condition at $x=L$): Eq. (\ref{nl3}) gives  
\begin{equation} 
e^{2i k L} = \left ({k+i \eta_i\over k-i \eta_i}\right ) \, ,   
\label{nl5}
\end{equation} 
which gives 
\begin{eqnarray} 
\eta_i &=& 0 \Longrightarrow  \phi_i(n,x) = \sqrt {{2\over L}} 
\cos \left [(n-1) \pi {x\over L}\right ] \, , \qquad n=1,2,...
\label{nl6}\\
\eta_i &=& {\infty} \Longrightarrow \phi_i(n,x) = 
\sqrt {{2\over L}}\sin \left [\left (n-{1\over 2}\right ) 
\pi {x\over L}\right ]\, ,  
\qquad n=1,2,... \, ; 
\label{nl7}
\end{eqnarray}
\medskip 

(b) $\mu =\infty $ (Dirichlet condition at $x=L$): Eq. (\ref{nl3}) implies 
\begin{equation} 
e^{2i k L} = -\left ({k+i \eta_i\over k-i \eta_i}\right ) \, ,   
\label{nl8}
\end{equation} 
which gives 
\begin{eqnarray} 
\eta_i &=& 0 \Longrightarrow  \phi_i(n,x) = \sqrt {{2\over L}} 
\cos \left [\left (n-{1\over 2}\right ) \pi {x\over L}\right ] \, , 
\qquad n=1,2,...
\label{nl9}\\
\eta_i &=& {\infty} \Longrightarrow \phi_i(n,x) = \sqrt {{2\over L}} 
\sin \left (n \pi {x\over L}\right ) \, , \qquad n=1,2,... 
\label{nl10}
\end{eqnarray} 
Notice that any of the sets (\ref{nl5}, \ref{nl6}, \ref{nl9}, \ref{nl10}) represent a complete 
ortho-normal system in ${\mathbb L}^2[0,L]$. 

\bigskip
\subsection{Scale invariant scattering matrices} 
\bigskip 

Observing that the eigenvalue of ${\mathbb S}$ is $1$ for $\eta_i=0$ and $-1$ for $\eta_i=\infty$ 
one concludes that the most general scale-invariant scattering matrix, compatible with a unitary time evolution, 
is given by 
\begin{equation} 
\S = \U\, \S_d\, \U^\dagger\, , \qquad \S_d = {\rm diag}(\pm 1, \pm 1, \cdots \pm 1)\, ,  
\label{sd}
\end{equation} 
where $\U$ is a generic $M\times M$ unitary matrix. From the group-theoretical point of view, 
any critical $\S$ matrix is a point in the orbit of some $\S_d$ under the {\it adjoint} action of the 
unitary group $U(M)$. Obviously, one can enumerate the 
edges in such a way that the first $p$ eigenvalues of $\S$ are $+1$ and the remaining $M-p$ are 
$-1$. The cases $p=M$ and $p=0$ correspond to $\S=\II$ and $\S=-\II$ and are 
not interesting for the entanglement. In these two cases in fact, the single wires 
are decoupled (there is no transmission), which implies a vanishing entanglement. 

It follows from (\ref{sd}) that besides being unitary, $\S$ is also Hermitian (in agreement 
with (\ref{S3}) at criticality). Therefore, all diagonal elements $\S_{ii}$ are real. Notice however that 
in general $\S$ is not symmetric. If this is the case, time reversal invariance is broken \cite{Bellazzini:2009nk}. 

In the nontrivial case $p=1$ the most general $2\times 2$ scale-invariant 
scattering matrix depends on two parameters and can be written in the form 
\begin{equation} 
{\mathbb S}(\epsilon, \theta) = 
{1\over 1+\epsilon^2}\left(\begin{array}{cc}\epsilon^2-1 & 2\epsilon
e^{i \theta}\\ 2\epsilon e^{-i \theta}& 1-\epsilon^2  
\\ \end{array} \right)\, , \qquad \epsilon \in \RR,\quad \theta \in [0,2\pi)\, .
\label{n21}
\end{equation} 
Time reversal invariance is broken for $\theta \not=0,\pi$. 

For $M=3$ one has two families corresponding to $p=1$ and $p=2$. In order to avoid 
cumbersome formulae, we display only two representatives of these families, namely  
\begin{equation} 
{\mathbb S}_{p=1}(\epsilon_1,\epsilon_2) ={1\over 1+\epsilon_1^2 +\epsilon_2^2} 
\left(\begin{array}{ccc}
2\epsilon_1&2\epsilon_2&1-\epsilon_1^2-\epsilon_2^2 \\
2\epsilon_2&-\epsilon_1^2 +\epsilon_2^2 -1&2\epsilon_1\epsilon_2\\
\epsilon_1^2-\epsilon_2^2 -1&2\epsilon_1 \epsilon_2 &2\epsilon_1
\end{array}\right) \, , 
\label{p1}
\end{equation} 
\begin{equation} 
{\mathbb S}_{p=2}(\epsilon_1,\epsilon_2) ={-1\over 1+\epsilon_1^2 +\epsilon_2^2} 
\left(\begin{array}{ccc}
\epsilon_1^2-\epsilon_2^2 -1&2\epsilon_1 \epsilon_2 &2\epsilon_1 \\ 
2\epsilon_1\epsilon_2&-\epsilon_1^2 +\epsilon_2^2 -1&2\epsilon_2\\
2\epsilon_1&2\epsilon_2&1-\epsilon_1^2-\epsilon_2^2  
\end{array}\right)\, , 
\label{p2}
\end{equation} 
where $\epsilon_{1,2} \in \RR$. 

\bigskip
\subsection{Two-point correlation function} 
\medskip

Now we are in position to construct the physical field $\psi_i(t,x)$ and the 
relative two-point function needed in the computation of the entanglement entropy. 
First of all, we write the unphysical field in terms of the eigenfunctions $\phi_i(n,x)$
\begin{equation}
\varphi_i(t,x) = \sum_{n=1}^\infty \e^{-\ri \omega_i(n)t}\, \phi_i(n,x) a_i(n) \, , 
\label{a1}
\end{equation} 
where the fermion annihilation and creation operators satisfy standard anti-commutation relations 
\begin{equation}
[a_i(m)\, ,\, a^\dagger_j(n)]_+=\delta_{ij} \delta_{mn}\, , \qquad 
[a_i(m)\, ,\, a_j(n)]_+=[a_i^\dagger(m)\, ,\, a^\dagger_j(n)]_+=0\, , 
\label{a3}
\end{equation}
and the energies are given by
\begin{equation}
\omega_i(n) = 
\begin{cases} 
\frac{1}{2m}\left [(n-1)\frac{\pi}{L}\right ]^2\, , & \qquad  \text{if\; $1\leq i\leq p$}\, , \\
\frac{1}{2m}\left [(2n-1)\frac{\pi}{2L}\right ]^2\, , & \qquad  \text{if\; $p<i\leq M$}\, , \\ 
\end{cases}
\label{d12}
\end{equation} 
\begin{equation}
\omega_i(n) = 
\begin{cases} 
\frac{1}{2m}\left [(2n-1)\frac{\pi}{2L}\right ]^2\, , & \qquad  \text{if\; $1\leq i\leq p$}\, , \\
\frac{1}{2m}\left [n\frac{\pi}{L}\right ]^2\, , & \qquad  \text{if\; $p<i\leq M$}\, , \\ 
\end{cases}
\label{d122}
\end{equation} 
for $\mu=0$ and $\mu=\infty$ respectively. Notice that different ``unphysical" edges 
may have different dispersion relation, which 
is not a problem because these edges are totally isolated from each other. 
By means of (\ref{nl1}) one gets the physical fields 
\begin{equation} 
\psi_i(t,x) = \sum_{j=1}^M \U^\dagger_{ij}\, \varphi_j(t,x) = 
\sum_{j=1}^M\sum_{n=1}^\infty  \U^\dagger_{ij}\, \e^{-\ri \omega_j(n)t}\, \phi_j(n,x) a_j(n) \, . 
\label{a5} 
\end{equation}  
One easily verifies that 
\begin{equation}
[\psi_i(t,x)\, ,\, \psi^\dagger_j(t,y)]_+ = \delta_{ij} \delta (x-y)\, , 
\label{b}
\end{equation}
which fixes the normalization of the fields. 

The equal time two-point correlation function of the physical field $\psi_i(t,x)$ on a given 
state $| \Psi \rangle$ is
\bea
C_{ij}^\Psi(x,y)&\equiv& 
\langle \Psi| \psi^\dagger_i(t,x) \psi_j(t,y)|\Psi \rangle 
\nonumber\\ &=& 
\sum_{k,l=1}^M\sum_{n,m=1}^\infty 
\U^\dagger_{jk}\, \U_{li} \e^{\ri [\omega_j(m)-\omega_i(n)]t} \phi_k(n,x)\overline{\phi}_l(m,y) 
\langle \Psi|  a^\dagger_k(n)a_l(m) | \Psi \rangle\, , 
\label{a7}
\eea
where the correlator $\langle \Psi|  a^\dagger_k(n)a_l(m) | \Psi \rangle$ can be deduced from the action of the 
algebra generated by $\{a_i(m)\, , a^\dagger_j(n)\}$ on the state $|\Psi\rangle$.
In particular, we are interested in the case when $|\Psi\rangle$ is the ground-state of the system 
formed by $N$ fermions in the whole junction. 
It is then useful to rewrite $N$ as 
\begin{equation}
N = M\, \N 
\label{pn}
\end{equation}
where $\N$ represents the average number of particles for each wire.
The action of the annihilation and creation operators on the ground-state is obvious since 
it is annihilated by all $a_l(m)$ with $m> \N$ and so 
$\langle \Psi|  a^\dagger_k(n)a_l(m) | \Psi \rangle=\delta_{kl} \delta_{nm} \theta(\N-n) $. 
Using this relation,
Eq.  (\ref{a7}),  restricted to the same edge ($i=j$) which is needed actually for computing the entanglement entropy, 
becomes
\begin{equation} 
C^N_{ii}(x,y) = \sum_{k=1}^M\sum_{n=1}^\N 
|\U_{ki}|^2 \, \phi_k(n,x)\overline{\phi}_k(n,y) \, , 
\label{a8}
\end{equation} 
where $\N$ can also be interpreted as an ultraviolet cut-off for the series in (\ref{a7}).

It is convenient for what follows to rewrite (\ref{a8}) in more explicit terms. For this purpose we 
consider any critical point characterized by the integer $1 < p < M$, i.e. a scale invariant 
scattering matrix with $p$ eigenvalues equal to $+1$. The two sums in (\ref{a8}) factorize and one gets 
\begin{equation} 
C^N_{ii}(x,y) = 
\sum_{k=1}^p |\U_{ki}|^2 \sum_{n=1}^\N f_+(n,x)f_+(n,y)
+ \sum_{k=p+1}^M |\U_{ki}|^2 
\sum_{n=1}^\N f_-(n,x)f_-(n,y)\, ,   
\label{a9}
\end{equation} 
where 
\begin{equation} 
f_+(n,x)= 
\begin{cases} 
\sqrt {{2\over L}} \cos \left [(n-1) \pi {x\over L}\right ]\, , & \qquad  \quad \; \mu = 0\, , \\ 
\sqrt {{2\over L}} \cos \left [\left (n-{1\over 2}\right ) \pi {x\over L}\right ]\, , & \qquad  \quad \; \mu=\infty \, , \\ 
\end{cases} 
\label{d1}
\end{equation}
\begin{equation}
f_-(n,x)= 
\begin{cases} 
\sqrt {{2\over L}}\sin \left [\left (n-{1\over 2}\right )\pi {x\over L}\right ]\, , & \qquad \qquad  \mu = 0\, , \\ 
\sqrt {{2\over L}}\sin \left [n \pi {x\over L}\right ]\, , & \qquad  \qquad \mu=\infty \, . \\ 
\end{cases}  
\label{d2}
\end{equation} 
Because of (\ref{sd}), the sums over $k$ in (\ref{a9}) give 
\begin{equation} 
\sum_{k=1}^p |\U_{ki}|^2 = \frac{1}{2} \left (1+\S_{ii}\right ) \, , 
\qquad 
\sum_{k=p+1}^M |\U_{ki}|^2 = \frac{1}{2} \left (1-\S_{ii}\right )\, . 
\label{c4}
\end{equation} 
Moreover, using that $\S$ is both unitary and Hermitian, one has 
\begin{equation}
\S_{ii}^2 = 1 - \sum_{j=1 \atop {j\not=i}}^M |\S_{ij}|^2 \equiv 1-T_i^2\, , 
\label{c5}
\end{equation}
where $T_i^2$ is the {\it total transmission probability} from the edge $i$ to the rest of the graph. 

In conclusion, the correlator (\ref{a8}) can be fully expressed in 
terms of the transmission probability $T^2_i$ 
and the one-particle wave functions as follows 
\begin{equation} 
C^N_{ii}(x,y) = 
\frac{1}{2} \left (1+\sqrt {1-T^2_i}\right ) \sum_{n=1}^\N f_+(x,n)f_+(y,n)
+ \frac{1}{2} \left (1-\sqrt {1-T^2_i}\right )
\sum_{n=1}^\N f_-(x,n)f_-(y,n)\, ,   
\label{a10}
\end{equation} 
which is the basic input for deriving the entanglement entropy in the next section. 
We observe that $C^N_{ii}$ involves only the diagonal elements of $\S$ and consequently,  
does not depend on the behavior of $\S$ under transposition. Therefore, 
contrary to the conductance \cite{Bellazzini:2009nk}, the 
entanglement entropy in our case is not sensitive to the breaking of time-reversal invariance. 

\bigskip 
\section{Entanglement entropy} 
\medskip 

In order to compute the bipartite R\'enyi entanglement entropies defined as in Eq. (\ref{Sndef}) of a subsystem $A$ 
in the ground-state our star graph, we use the method recently introduced in Refs. \cite{Calabrese:2011zz, Calabrese:2011vh}. 
The starting point to deal with a system made of a finite number of particles in continuous space is the  Fredholm determinant
\begin{equation}
\D_A(\lambda) =
{\rm det}\left[\lambda \delta_A(x,y)- C_A(x,y)\right] \,,
\label{dl}
\end{equation}
where $C_A(x,y)$ is the restriction of the correlation matrix $C(x,y)$ defined in Eq. (\ref{a7}) to $A$, 
i.e. $C_A=P_A C P_A$, where $P_A$ is the projector on $A$.  
The same definition holds for $\delta_A(x,y)=P_A\delta(x-y)P_A$.  
Following the ideas for the lattice model \cite{jk-04}, $\D_A(\lambda)$ can be introduced in
such a way that it is a polynomial in $\lambda$ having as zeros the eigenvalues of $C_A$.  
Since we are dealing only with free fermions in the bulk,  the reduced density matrix $\rho_A$ is Gaussian \cite{ep-rev}
and so one can easily derive \cite{jk-04,Calabrese:2011zz, Calabrese:2011vh}
\begin{equation}
S^{(\a)} \equiv \frac{\ln {\rm Tr}\rho_A^\a}{1-\a} = 
 \oint \frac{d \lambda}{2\pi i}\, e_\a(\lambda) 
\frac{d \ln \D_A(\lambda)}{d\lambda},
\label{snx}
\end{equation}
where the integration contour encircles the segment $[0,1]$, and
\begin{equation}
e_\a(\lambda) = {1\over 1-\a} 
\ln \left[{\lambda}^\a
+\left({1-\lambda}\right)^\a\right]\,.
\label{enx}
\end{equation}
For $\a\to 1$,  $e_1(\lambda)=-x\ln x-(1-x)\ln(1-x)$ and Eq. (\ref{snx}) gives the von Neumann entropy.  

The Fredholm determinant is turned into a standard one by introducing
the {\em reduced overlap} matrix ${\mathbb A}$ (also considered in
Ref.~\cite{Klich-06}) with elements
\begin{equation}
{\mathbb A}_{nm} =  \int_{x_1}^{x_2} dz\, \overline{\phi}_n(z) \phi_m(z),
\qquad n,m=1,...,D,
\label{aiodef}
\end{equation}
where in general $\phi_n(x)$ represent the eigenfuctions corresponding to the $D$ lowest 
energy level which are occupied in the ground-state of the system with $D$ degrees of freedom.
The matrix ${\mathbb A}$ satisfies $ {\rm Tr}\, C_A^k = {\rm Tr}\,{\mathbb A}^k$ and so \cite{Calabrese:2011vh}
\be
\ln {{\mathbb D}}_A(\lambda) =
-\sum_{k=1}^\infty { {\rm Tr} C_A^k \over k \lambda^k }=
-\sum_{k=1}^\infty { {\rm Tr} {\mathbb A}^k \over k \lambda^k }=
\ln  {\rm det}\left[{\lambda {\mathbb I}-{{\mathbb A}}}\right] = 
\sum_{m=1}^D \ln(\lambda-a_m)\,, 
\label{logdetn}
\ee 
where $a_m$ are the eigenvalues of ${\mathbb A}$ and $D$ is its dimension to be specified later. 
Inserting (\ref{logdetn}) in the integral (\ref{snx}), we obtain
\begin{equation}
S^{(\a)} =  \oint  \frac{d \lambda}{2\pi i}
\sum_{m=1}^D {e_\a(\lambda)\over  \lambda  - a_m}  = 
\sum_{m=1}^D e_\a(a_m) =  \frac{1}{1-\alpha} {\rm Tr}\ln \left [\A^\alpha + (\II-\A)^\alpha \right ] \, ,
\label{snx2n}
\end{equation}
as a consequence of the residue theorem.

In the following we will be interested only in the 
entanglement entropy of any edge $i$ of the wire with respect to the rest of the junction in the global ground-state 
of the star graph. 
As we have seen above in Eq. (\ref{a10}), the two-point correlation function for finite number of particles $N$ in the full star graph, 
can be written in the form (we omit the edge index $i$ hereafter) 
\begin{equation}
C^N(x,y) = \sum_{n=1}^{2\N} {\overline \chi}(x,n) \chi (y,n) \, , 
\label{e1}
\end{equation} 
where $\chi(x,n)$ are proportional to the one-particle eigenfunctions $f_\pm(x,n)$ in Eq. (\ref{a10}). 
Then the R\'enyi entanglement entropy of the subsystem represented by a single wire is given by 
Eq. (\ref{snx2n}) where the eigenvalues $a_m$ of ${\mathbb A}$ are  numerically calculated
from the overlap matrix built with the correlation function above. 

\bigskip 
\subsection{The case $M=2$} 
\medskip 

It is instructive to consider first the case with two edges only. In this case, one has actually a segment of length $2L$ 
with a point-like conformal defect placed in the middle. This situation has been investigated on the lattice
\cite{ep-10,p-11} by other methods and represents a useful check for our framework. 
We set $L=1$ for simplicity and consider the case $\mu=\infty$ 
(Dirichlet condition at $x=L$).\footnote{The case $\mu=0$ (Neumann boundary 
conditions at $x=L$) can be treated along the same lines.}
Combining the explicit form (\ref{n21}) of the $\S$-matrix 
with (\ref{c5}), one obtains the following transmission amplitudess 
\begin{equation}
T_1 = T_2 = \frac{2\epsilon}{1+\epsilon^2} \equiv T\, ,
\label{t12}
\end{equation}
which are the square root of the transmission probability. 
Accordingly, 
\begin{equation}
C^N_{11}(x,y) = C^N_{22}(x,y) = \sum_{n=1}^{2\N} {\overline \chi}(x,n) \chi (y,n) \, , 
\label{e5}
\end{equation} 
with 
\begin{equation}
\chi (k,x)= 
\begin{cases} 
\frac{\epsilon}{\sqrt {{1+\epsilon^2}}}\sqrt {2} \cos \left (k\pi {x\over 2}\right )\, , & \qquad k=1,3,...,2\N-1\, , \\ 
\frac{1}{\sqrt {{1+\epsilon^2}}}\sqrt{2} \sin \left (k\pi {x\over 2}\right )\, , & \qquad k=2,4,...,2\N\, , \\ 
\end{cases}  
\label{e6}
\end{equation} 
where $N=2\N$ is the total particle number according to eq. (\ref{pn}). 
The matrix $\A$, defined by (\ref{aiodef}), reads in our case 
\begin{eqnarray} 
\A_{mn} &=& \frac{\epsilon^2}{(1+\epsilon^2)} \delta_{mn}\, , \qquad \qquad \quad \; m,n - {\rm odd}\, , 
\label{m1}\\ 
\A_{mn} &=& \frac{1}{(1+\epsilon^2)} \delta_{mn}\, , \qquad \qquad \quad \; m,n - {\rm even}\, , 
\label{m2}\\ 
\A_{mn} &=& \frac{2\epsilon}{(1+\epsilon^2)} \frac{2n}{\pi(n^2-m^2)} \, , \qquad m - {\rm odd}\, , n - {\rm even}\, , 
\label{m3}\\ 
\A_{mn} &=& \frac{2\epsilon}{(1+\epsilon^2)} \frac{2m}{\pi(m^2-n^2)} \, , \qquad m - {\rm even}\, , n - {\rm odd}\, . 
\label{m4} 
\end{eqnarray} 
Since $C^N_{11} = C^N_{22}$, the entanglement entropy of the edge 1 equals that of the edge 2 and 
is given by 
\begin{equation} 
S^{(\a)}(T;N) = \frac{1}{1-\alpha} {\rm Tr}\ln \left [\A^\alpha + (\II-\A)^\alpha \right ]  = \sum_{n=1}^{N} e_\alpha (a_n) \, , 
\label{e7}
\end{equation}
$a_n$ being the eigenvalues of the matrix (\ref{m1}-\ref{m4}). Using (\ref{e7}), we will show below that 
\begin{equation}
S^{(\a)} (T;N) = {\cal C}^{(\alpha )}(T) \ln N + O(1) \, ,
\label{asbehb}
\end{equation} 
with a pre-factor $\C^{(\alpha)}(T)$ that depends on the transmission 
amplitude and not only on the central charge $c=1$. 

Before considering a generic value of $\alpha$, it is instructive to discuss the R\'enyi entropy $\alpha =2$. 
In this case Eq. (\ref{e7}) takes the simple form
\begin{equation} 
S^{(2)}(T;N) =  - {\rm Tr}\, \ln \left [\II - 2 \E (T) \right ] =  
\sum_{k=1}^\infty \frac{2^k}{k}\, {\rm Tr}\, \E^k(T) \, , 
\label{n1}
\end{equation}
where the combination 
\begin{equation}
\E (T) \equiv \A (\II-\A)   
\label{n2}
\end{equation} 
represents the natural variable for performing the computation. 
The main idea at this point is to reduce the evaluation of (\ref{n1}) to the case of full transmission $T=1$, 
when the defect is absent and one can use therefore the known \cite{cmv-11} behavior of ${\rm Tr}\, \E^k$
for free fermion gas on the interval, namely 
\begin{equation}
{\rm Tr}\, \E^k(T=1) = \frac{1}{2\pi^2} \frac{[(k-1)!]^2}{(2k-1)!} \ln N + O(1) \, . 
\label{n3}
\end{equation}
{}For this purpose we first establish the fundamental relation  
\begin{equation}
{\rm Tr}\, \E^k(T) = T^{2k}\, {\rm Tr}\, \E^k(T=1)\, , 
\label{n4} 
\end{equation}
which captures the impact of the point-like interaction in the junction on the entanglement entropy (\ref{n1}). 
In order to prove (\ref{n4}), we exploit the property that a reordering of rows and columns of $\E$ 
does not change the trace of $\E^k$, we are interested in.
We then write the overlap matrix in the following block form 
\begin{equation} 
\A = \left(\begin{array}{cc}\frac{\epsilon^2}{1+\epsilon^2} \II & T\, \B_1
\\T\, \B_2 & \frac{1}{1+\epsilon^2}\II
\\ \end{array} \right) \,,
\label{n5} 
\end{equation}
with quadratic blocks of size $\N$. Here $\II$ is the $\N\times \N$ identity matrix and 
with respect to (\ref{m1}-\ref{m4}) we have reordered the lines and rows of 
$\A$ in such a way that the upper block on the left and the lower one on the right 
coincide with (\ref{m1}) and (\ref{m2}) respectively (i.e. after this reordering the first lines and rows 
are the odd indices $n,m$, while the right-lower block is formed by even $n,m$). 
$\B_1$ and $\B_2$ are $T$-independent matrices, whose explicit form is not essential for the proof. 
Using the representation (\ref{n5}) and the relation (\ref{t12}), one gets 
\begin{equation}
\E = \A (\II - \A) = 
\left(\begin{array}{cc}\frac{\epsilon^2}{1+\epsilon^2}\, \II & T\, \B_1
\\T\, \B_2 & \frac{\epsilon^2}{1+\epsilon^2}\, \II
\\ \end{array} \right) 
\left(\begin{array}{cc}\frac{1}{1+\epsilon^2} \II & -T\, \B_1
\\-T\, \B_2 & \frac{\epsilon^2}{1+\epsilon^2}\II
\\ \end{array} \right) = \frac{T^2}{4} \left(\begin{array}{cc}\II +\B_1\, \B_2 & 0 
\\0  & \II + \B_2\, \B_1 
\\ \end{array} \right)\, , 
\label{n6}
\end{equation}
which proves (\ref{n4}). Finally, plugging (\ref{n3},\ref{n4}) in (\ref{n1}) one obtains 
\begin{equation} 
\C^{(2)}(T) =   
\sum_{k=1}^\infty \frac{2^{k-1}[(k-1)!]^2}{\pi^2k(2k-1)!} T^{2k} = \frac{2}{\pi^2} \arcsin^2\left (\frac{T}{\sqrt 2}\right )\, ,  
\label{n7}
\end{equation}
where we used 
\be
\sum_{k=1}^\infty \frac{[(k-1)!]^2}{(2k)!} (2x)^{2k}=2\arcsin^2x \,.
\label{sumsin}
\ee
We stress that each integer value of $\alpha \geq 2$ can be treated in 
analogous way. For $\alpha = 3, 4$ one finds for instance 
\begin{equation} 
\C^{(3)}(T) =   
\sum_{k=1}^\infty \frac{3^k[(k-1)!]^2}{\pi^2 2 k (2k-1)!} T^{2k} = \frac{2}{\pi^2} \arcsin^2\left (\frac{T\sqrt {3}}{2}\right )\, ,  
\label{n8}
\end{equation}
and 
\begin{eqnarray} 
\C^{(4)}(T) =   
\sum_{k=1}^\infty \frac{[(2+\sqrt {2})^k +(2-\sqrt {2})^k] 
[(k-1)!]^2}{6 \pi^2k(2k-1)!} T^{2k} = 
\nonumber \\
\frac{2}{3\pi^2} \left [\arcsin^2\left (\frac{T}{2}\sqrt {2+\sqrt{2}}\,\right ) + 
\arcsin^2\left (\frac{T}{2}\sqrt {2-\sqrt{2}}\, \right )\right ]\, .  
\label{n9}
\end{eqnarray}

\subsubsection{Result for generic integer $\alpha$}

In order to obtain the result for generic integer $\alpha$ we first need to write the combination $\A^\alpha +(\II - \A)^\alpha$
in terms of the matrix $\E$. 
This can be achieved by formally inverting Eq. (\ref{n2}) as
\be
\A=\frac12(1\pm \sqrt{1-4\E})\,,
\ee
with an ambiguity in the choice of the sign reflecting the degeneration of the spectrum of $\E$.
However, in the needed combination
\be
\A^\alpha +(\II - \A)^\alpha=2^{-\alpha}\left[
(1\pm \sqrt{1-4\E})^\alpha+(1\mp \sqrt{1-4\E})^\alpha\right]\,,
\ee
the choice of this sign is unimportant. 
Notice that although the apparent presence of a square-root, for integer $\a$ the above expression is 
a polynomial in $\E$ of degree $\lfloor \alpha/2 \rfloor$, that is the integer part of $\alpha/2$.
Using the binomial theorem $(1+x)^\alpha=\sum_{k=0}^\alpha \binom{n}{k} x^k$, we have
\bea
\A^\alpha +(\II - \A)^\alpha&=&2^{-\alpha}\sum_{k=0}^\alpha\binom{\a}{k} (1-4\E)^{k/2} (1-(-1)^k)\\ &=&
2^{1-\alpha}\sum_{k=0}^{\lfloor \alpha/2 \rfloor}\binom{\a}{2k} (1-4\E)^k\,,
\eea
which makes the polynomial form explicit.
Using again the binomial theorem for $(1-4\E)^k$, we have 
\bea
\A^\alpha +(\II - \A)^\alpha&=&
2^{1-\alpha}\sum_{k=0}^{\lfloor \alpha/2 \rfloor}\binom{\a}{2k} \sum_{p=0}^k\binom{k}{p} (-4\E)^k=
-\sum_{p=0}^{\lfloor \alpha/2 \rfloor} v_p (-4\E)^p
\,,\nonumber
\eea
where we defined
\be
v_p\equiv - 2^{1-\alpha}\sum_{k=p}^{\lfloor \alpha/2 \rfloor}\binom{\a}{2k} \binom{k}{p}
=- \binom{\a}{2p}\frac{\Gamma(\a-p) \Gamma(p+1/2)}{\sqrt{\pi} \Gamma(\a)}\,.
\ee

In order to calculate the R\'enyi entropies, we expand in series of $\E$ the quantity\break $\ln [\A^\a+(1-\A)^\a]$, 
obtaining 
\be
\ln [\A^\a+(1-\A)^\a]=\ln[1-\sum_{p=1}^{\lfloor \alpha/2 \rfloor} v_p (-4\E)^p]=
-\sum_{j=1}^\infty \frac1j\Big[\sum_{p=1}^{\lfloor \alpha/2 \rfloor} v_p (-4\E)^p\Big]^j\,. 
\ee
Using now the multinomial identity we have
\be
\ln [\A^\a+(1-\A)^\a]=- \sum_{j=1}^\infty\frac1j {\sum_{k_i}}'
\frac{j!}{k_1!\dots k_{\lfloor \alpha/2\rfloor}!} \prod_{p=1}^{\lfloor \alpha/2\rfloor} [v_p(-4\E)^p]^{k_p}
\ee
where we introduced the symbol ${\sum'_{k_i}}$ for the constrained sum $ \sum_{k_1,\dots, k_{\lfloor \alpha/2\rfloor}}$ 
 with $\sum k_i=j$.
We now introduce a sum over $K$ which will be equal to $\sum p k_p$ with the help of a Kronecker 
delta:
\be
\ln [\A^\a+(1-\A)^\a]=-
\sum_{K=1}^\infty\sum_{j=1}^\infty{\sum_{k_i}}'
\frac{(j-1)!}{k_1!\dots k_{\lfloor \alpha/2\rfloor}!} (-4\E)^K \delta_{K,\sum_p p k_p} \prod_{p=1}^{\lfloor \alpha/2\rfloor} v_p^{k_p}\,.
\ee
Using the contour integral representation of the Kronecker delta over the unitary circle $|z|=1$
\be
\delta_{a,b} =\frac1{2\pi i}\oint dz z^{a-b-1}\,,
\ee
 we have that the above expression equals
\bea
&&\frac1{2\pi i}\oint d z \sum_{K=1}^\infty (-4\E)^{K} z^{ K-1} \sum_{j=1}^\infty \frac1j {\sum_{k_i}}'
\frac{j!}{k_1!\dots k_{\lfloor \alpha/2\rfloor}!}  \prod_{p=1}^{\lfloor \alpha/2\rfloor} (v_p z^{- p})^{k_p} \nonumber\\&&= \frac1{2\pi i}
 \sum_{K=1}^\infty (-4\E)^{K}  \oint d z z^{ K-1} 
 \ln (1+ \sum_{p=1}^{\lfloor \alpha/2 \rfloor} v_p z^{-p} )  \nonumber\\&&= \frac1{2\pi i}
 \sum_{K=1}^\infty (-4\E)^{K}  \oint d z z^{ K-1} \ln  \left[\frac{{(1+ \sqrt{1+z^{-1}})}^\a+{(1- \sqrt{1+z^{-1}})}^\alpha}{2^{\a}}\right],
\eea
where, in the last line, we recognized the expansion of the function from which we started from for a complex argument (no matrices).

The integral over $z$ can be performed with standard techniques of integrals on the complex plane. 
The various contributions come from the discontinuities at the cuts which are placed between the 
zeros of the argument of the logarithm, i.e. at the $z_p$ satisfying 
\be
{\left(1+ \sqrt{1+z_p^{-1}}\right)}^\a+{\left(1- \sqrt{1+z_p^{-1}}\right)}^\alpha=0\,.
\ee
All the solutions of this equation are simply found as
\be
z_p=
-\cos^2 \frac{\pi (2p-1)}{2\a}, \quad {\rm with}\;\; p=1,\dots, \a/2\,.
\ee
Thus the integral is given by 
\be
 \frac1{2\pi i} \oint d z z^{ K-1} \ln  \left[\frac{{(1+ \sqrt{1+z^{-1}})}^\a+{(1- (\sqrt{1+z^{-1}})}^\alpha}{2^{\a}}\right]=
 \sum_{p=1}^{\lfloor \alpha/2 \rfloor} \frac{z_p^K}K\,,
\ee 
implying
\be
\ln [\A^\a+(1-\A)^\a]=
-\sum_{K=1}^\infty (-4\E)^K   \frac{(-1)^{K}}{ K} \sum_{p=1}^{\lfloor \alpha/2 \rfloor} \cos^{2K} \left( \pi \frac{2p-1}{2\a}\right)\,.
\ee
In this final form, it is straightforward to take the trace using Eq. (\ref{n4}) to obtain
\be
{\rm Tr}\ln [\A^\a+(1-\A)^\a]= -
\frac1{2\pi^2} \ln N \sum_{K=1}^\infty T^{2K} 4^K
\frac{[(K-1)!]^2}{(2K-1)!}  \sum_{p=1}^{\lfloor \alpha/2 \rfloor} \cos^{2K} \left( \pi \frac{2p-1}{2\a}\right)\,.
\ee
Inverting the order of the sums, the sum over $K$ can be now performed using Eq. (\ref{sumsin}) and we have
\be
{\rm Tr}\ln [\A^\a+(1-\A)^\a]=-\ln N \frac{2}{\pi^2}
\sum_{p=1}^{\lfloor \alpha/2 \rfloor} 
 \arcsin^2 \left [T\cos \frac{(2p-1)\pi}{2\alpha}\right]\,, 
\ee
that leads to the coefficient
\be
{\cal C}^{(\alpha )}(T)=\frac1{\a-1}\frac{2}{\pi^2}
\sum_{p=1}^{\lfloor \alpha/2 \rfloor} 
 \arcsin^2 \left[T\cos \frac{(2p-1)\pi}{2\alpha}\right]\,.
 \label{Cal-final}
\ee
For $\a=2,3,4$ it coincides with the result reported in the previous subsection. 

We observe that ${\cal C}^{(\a)}(0)=0$ and ${\cal C}^{(\a)}(1)$ provide useful checks.
 The value 
$T=0$ describes a totally reflecting defect; the two edges are isolated and indeed the 
entanglement vanishes.
For $T=1$ we have
\be
{\cal C}^{(\alpha )}(T=1)=\frac1{\a-1}\frac{2}{\pi^2}
\sum_{p=1}^{\lfloor \alpha/2 \rfloor} 
 \left[\frac\pi2- \frac{(2p-1)\pi}{2\alpha}\right]^2= \frac1{12}\left(1+\frac1\a\right)\,.
\ee
This corresponds to full transmission, i.e. the defect is absent and one is considering the 
entanglement entropy of half of a system of length $2L$ with boundaries obtained in \cite{Calabrese:2011zz} 
and compatible with the standard CFT result \cite{cc-04}.

\subsubsection{The analytic continuation}

In order to investigate non-integer values of $\alpha$, we exploit the
results of Ref. \cite{ep-10,p-11} for the Ising and XX  spin-chains. 
Using methods based on corner transfer matrix and conformal mappings, Eisler and Peschel  
derived the entanglement entropy of the subsystem on (let us say) the right of the defect, for 
a chain of length $2L$ with a defect in the middle.
For the $XX$ chain, that after a Jordan-Wigner transformation corresponds to a lattice gas of free spinless fermions, 
the result can be written in the form \cite{ep-10}
\be
S^{(\a)} (T;L) = {\cal C}^{(\alpha )}_L(T) \ln L + O(1)  \,,
\label{ep}
\ee
with ${\cal C}^{(\alpha )}_L(T)$ given by
\begin{equation}
\C_L^{(\alpha )}(T) = \frac{2}{\pi^2 (1-\alpha)} 
\int_0^\infty \rd x \ln \left [\frac{1+\e^{-2 \alpha \omega(x,T)}}
{\left(1+\e^{-2\omega(x,T)}\right )^\alpha }\right ] \, , 
\label{e8}
\end{equation}  
where 
\begin{equation} 
\omega (x,T) = {\rm acosh} \left [\frac {\cosh (x)}{T}\right ] \, . 
\label{e9}
\end{equation}  
Actually only the result for $\a=1$ has been reported in Ref. \cite{ep-10}, but the derivation for general $\a$ is straightforward from 
the results reported there. 
A result of $\a=2$ has been reported in \cite{p-11} and for general integer $\a$ in \cite{p-pc}.
For a simple comparison between our work and Ref. \cite{ep-10}, we mention that the transmission amplitude $T$ in Ref. 
\cite{ep-10} is called $s$.   

For integer $\a$, it is straightforward to check numerically that $\C_L^{(\alpha )}(T)$ and $\C^{(\alpha )}(T) $ in Eq. (\ref{Cal-final}) 
are equal, as it is possible to show analytically \cite{p-pc}. 
This coincidence does not come unexpected. Indeed, 
since Eq. (\ref{ep}) is valid for finite density $N/2L=1/2$ on the lattice, it can be turned in 
the entanglement entropy as function of $N$, simply by replacing $L$ with $N$, but
assuming (as we did in \cite{Calabrese:2011zz} on the basis of numerical data) that the dependence on $T$ is universal.
Taking now the continuum limit, we straightforwardly deduce that 
\be
{\cal C}^{(\a)}(T)={\cal C}_L^{(\a)}(T)\,.
\ee
However, the above computation proves the universal $T$ dependence with no assumption. 
The computation of Ref. \cite{p-pc} also shows that the result in Eq. (\ref{e8}) is the 
analytic continuation of Eq. (\ref{Cal-final}) to non-integer values of $\a$. 
In particular, in the von Neumann case ($\a=1$), the integral can be performed and one has \cite{ep-10}
\begin{equation}
{\cal C}_L^{(1)}(T) = {1\over \pi^2} \left\{\left[ (1+T)\ln(1+T)+ (1-T)\ln(1-T)\right]
\ln T + (1+T){\rm Li}_2(-T) + (1-T){\rm Li}_2(T)\right\} .
\label{bomega}
\end{equation}

\subsubsection{Comparison with numerical computation}
 
We now turn to briefly present the numerical data for the entanglement entropies 
which have been fundamental for the conceptual understanding that led to the 
exact computation in the previous subsection. 
Furthermore numerical computations give non trivial insights about the 
corrections to the asymptotic behavior.
In particular we anticipate that in analogy to what found for the half-space R\'enyi 
entanglement entropies in homogeneous systems \cite{Calabrese:2011vh,corrlatt}, the leading suppressed 
corrections turns out to be $O(N^{-1/\alpha})$.

\begin{figure}[t]
\begin{center}
\includegraphics[scale=0.3]{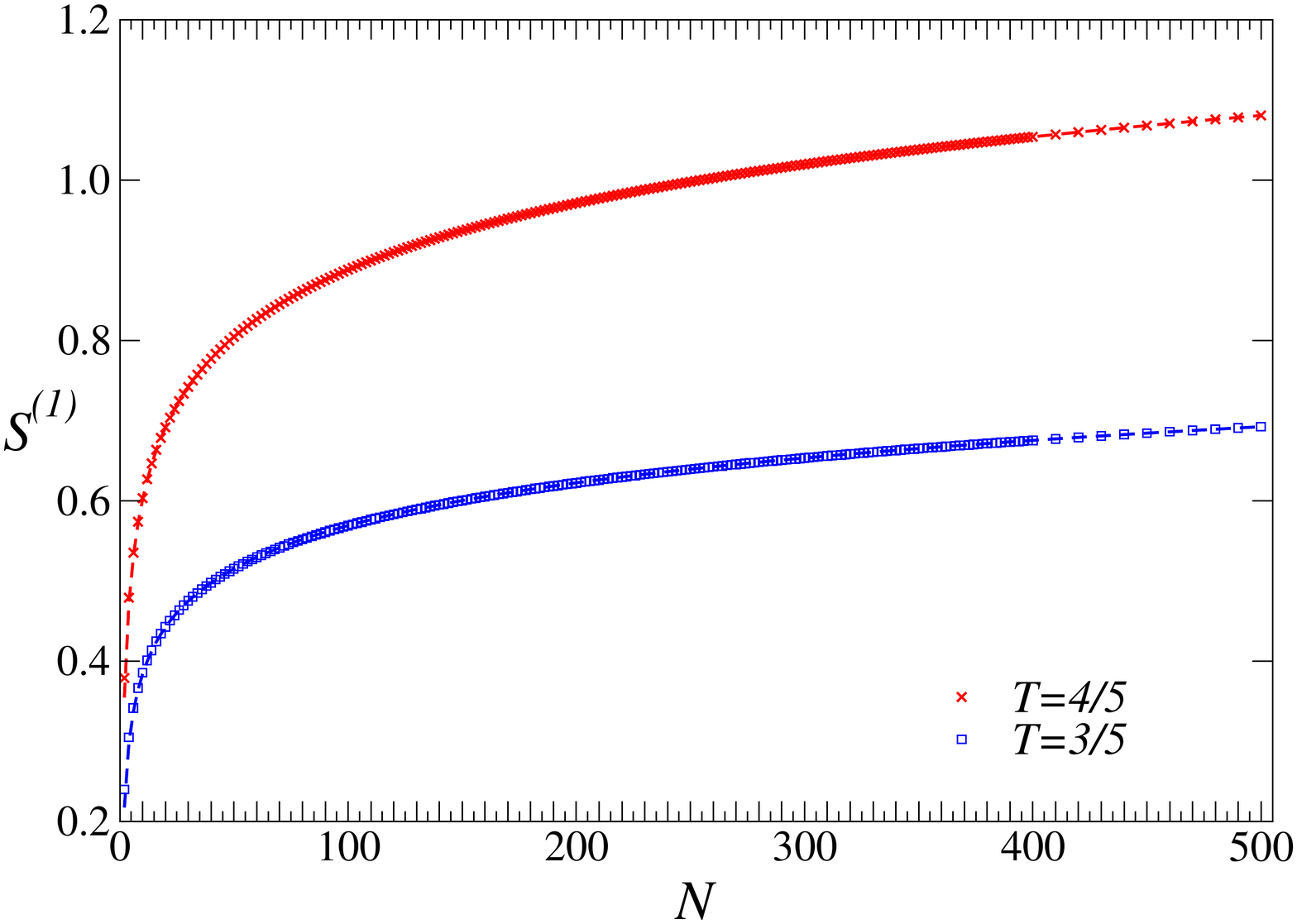}
\includegraphics[scale=0.3]{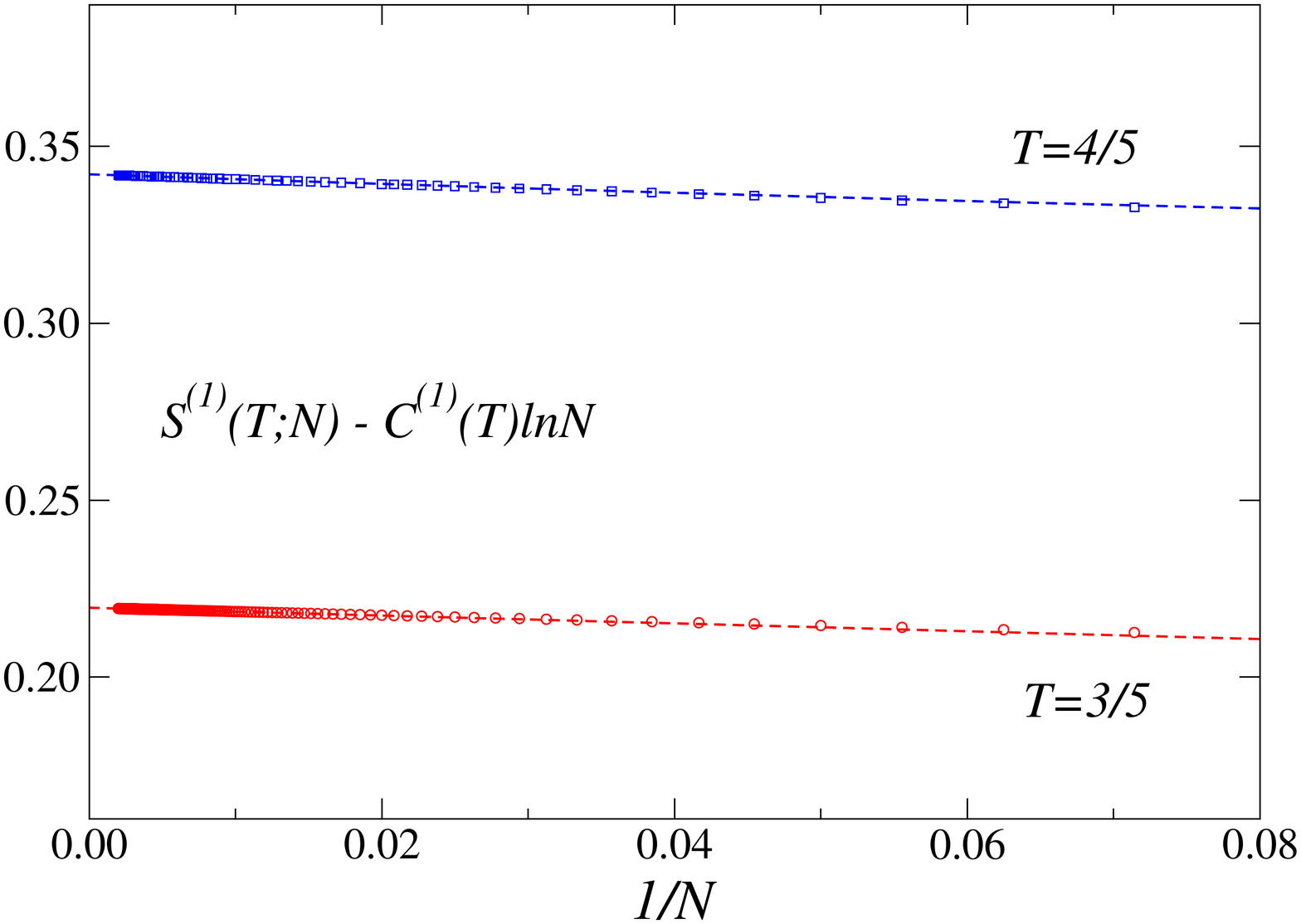}
\end{center}
\caption{(Color online)
The von Neumann edge entanglement entropy in the
two-wire junction for $T=3/5$ and $T=4/5$ where we considered only even values of $N$.  
Left:
We plot $S^{(1)}(T,N)$ vs $N$. 
Right:
The subtracted quantity $S^{(1)}(T,N) - {\cal C}^{(1)}(T)\ln N$, where ${\cal C}^{(1)}(T)$ is given by
Eq.~(\ref{e8}).  The lines
show fits of the data for $N\gtrsim 400$, where the correction to the
leading behavior ${\cal C}^{(1)}(T)\ln N$ is a polynomial
$b_0+b_1/N+b_2/N^2$.}\label{hbn2}
\end{figure}

The numerical estimates of the factor ${\cal C}^{(1)}(T)$, obtained
from our results for the entanglement entropy up to $N\approx 500$,
perfectly match the function (\ref{bomega}), within a precision better
than $O(10^{-6})$. Fig.~\ref{hbn2} shows the results for $T=3/5$ and
$T=4/5$ for even $N$. Fits to the $T=4/5$ data for $400\lesssim N \le 500$ 
\begin{equation}
S^{(1)}(T;N) = a \ln N + b_0 + b_1/N + b_2/N^2 + b_3/N^3
\label{fitans}
\end{equation}
give $a=0.118841$, $b_0=0.342056$, $b_1=-0.1404$, etc..., where $a$ should be compared
with ${\cal C}^{(1)}(4/5)=0.11884065...$. Analogous results are obtained
for $T=3/5$; we find $a=0.076078$ to be compared with ${\cal C}^{(1)}(3/5)=0.07607750$. 

\begin{figure}[t]
\begin{center}
\includegraphics[scale=0.4]{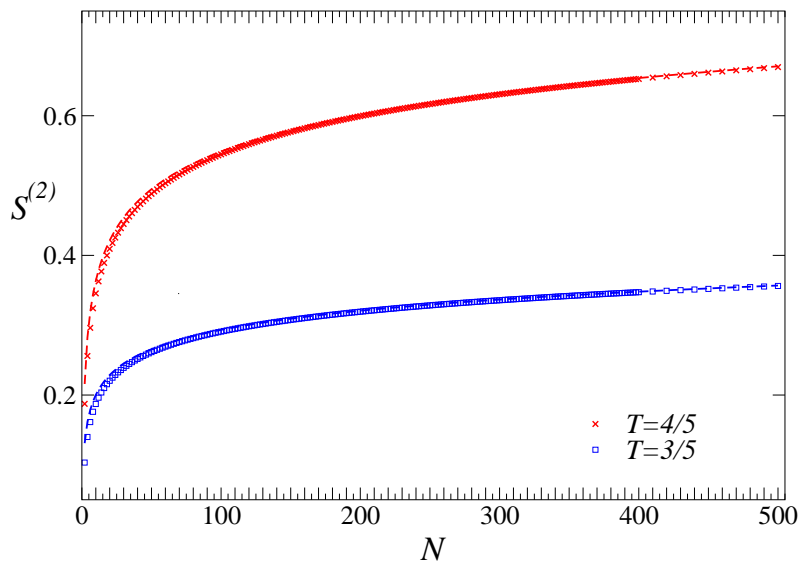}
\end{center}
\caption{(Color online) The $\alpha=2$ R\'enyi edge entanglement entropy 
$S^{(2)}(T,N)$ in the
two-wire junction for $T=3/5$ and $T=4/5$ for even number of particles $N$.  
The line show the curve
${\cal C}^{(2)}(T)\ln N + b_0 + b_1/N^{1/2} + b_2/N$ where the coefficients
$b_i$ are fitted using the data for $N\gtrsim 400$.}
\label{hbn2p2}
\end{figure}

A complete agreement is also found for the R\'enyi entropies.  For
$\alpha =2$ for instance, the data fit the asymptotic behavior
\begin{equation}
S^{(2)} (T;N) = {\cal C}^{(2)}(T) \ln N + b_0 + b_1/N^{1/2} + b_2/N + b_3/N^{3/2} + \cdots \, .
\label{fitans2}
\end{equation}
where the leading coefficient ${\cal C}^{(2)}(T)$ is given by 
Eq.~(\ref{n7}), as shown by Fig.~\ref{hbn2p2}.

In both Eqs. (\ref{fitans}) and (\ref{fitans2}) we use for the subleading corrections the form  $O(N^{-1/\alpha})$.
We mention that considering also odd values of the particles number $N$, we observe also in the 
graph entanglement entropies corrections to the scaling which depends on the parity of $N$,
in analogy to what observed in the absence of defects both in the continuum \cite{Calabrese:2011vh} and on the 
lattice \cite{corrlatt}.

\bigskip 
\subsection{The case of $M>2$ edges}
\medskip

The generalization to $M>2$ edges is now straightforward. The novelty is that now 
the entanglement entropy on the edge $i$ with respect to the whole junction depends on $i$. 
As expected from (\ref{a10}), this dependence is encoded in the transmission amplitudes $T_i$, which 
for $M>2$ do not coincide in general for different $i$. In order to investigate this aspect, it 
is convenient to introduce the variable 
\begin{equation} 
\Upsilon_i^2 = \frac{1}{2}\left (1+\sqrt{1-T^2_i}\right ) \, . 
\label{v1}
\end{equation} 
The two point-function (\ref{a10}) takes now the form 
\begin{equation} 
C^N_{ii}(x,y) = 
\Upsilon_i^2 \sum_{n=1}^\N f_+(x,n)f_+(y,n)
+ (1-\Upsilon_i^2)
\sum_{n=1}^\N f_-(x,n)f_-(y,n)\, .   
\label{v2}
\end{equation} 
Equivalently, using (\ref{d1},\ref{d2}), one finds in the case $\mu=\infty$ 
\begin{equation}
C^N_{ii}(x,y) = \sum_{n=1}^{2\N} {\overline \chi_i}(x,n) \chi_i (y,n) \, , 
\label{v3}
\end{equation} 
with
\begin{equation}
\chi_i (k,x)= 
\begin{cases} 
\Upsilon_i \sqrt {2}\cos \left (k\pi {x\over 2}\right )\, , & \qquad k=1,3,...,2\N-1\, , \\ 
\sqrt{1-\Upsilon_i^2} \sqrt {2}\sin \left (k\pi {x\over 2}\right )\, , & \qquad k=2,4,...,2\N\, . \\ 
\end{cases}  
\label{v4}
\end{equation} 
The reduced overlap matrix $\A$ now also carries an edge index $i=1,...,M$ and is given by 
\begin{eqnarray} 
\A^{(i)}_{mn} &=& 2 \Upsilon_i^2\, \delta_{mn}\, , \qquad \qquad \qquad \quad \quad \; \; m,n - {\rm odd}\, , 
\label{vm1}\\ 
\A^{(i)}_{mn} &=& 2(1-\Upsilon_i^2) \delta_{mn}\, , \qquad \qquad \qquad \; \, m,n - {\rm even}\, , 
\label{vm2}\\ 
\A^{(i)}_{mn} &=& 2\Upsilon_i \sqrt{1-\Upsilon_i^2} \frac{2n}{\pi(n^2-m^2)} \, , \qquad m - {\rm odd}\, , n - {\rm even}\, , 
\label{vm3}\\ 
\A^{(i)}_{mn} &=&  2\Upsilon_i \sqrt{1-\Upsilon_i^2} \frac{2m}{\pi(m^2-n^2)} \, , \qquad m - {\rm even}\, , n - {\rm odd}\, . 
\label{vm4} 
\end{eqnarray} 
Comparing the matrices (\ref{m1}-\ref{m4}) and (\ref{vm1}-\ref{vm4}) one concludes that 
the entanglement entropy $S_i^{(\alpha)}$ of the edge $i$ with respect to the whole junction 
depends on the edge via the transmission amplitude 
\begin{equation} 
T_i = 2 \Upsilon_i\, \sqrt{1-\Upsilon_i^2} \equiv \sqrt{1-\S_{ii}^2}  
\label{v5}
\end{equation}
and can be expressed by 
\begin{equation} 
S_i^{(\alpha)}(T_i;N) =  S^{(\alpha)}(T_i;N)\, , 
\label{v6}
\end{equation}
where $S^{(\alpha)}$ is the entropy (\ref{e7}) in the case $M=2$. Therefore, the asymptotic behavior 
for large $N$ is given by  
\begin{equation} 
S_i^{(\alpha)}(T_i;N) = {\cal C}^{(\alpha )}(T_i) \ln N + O(1) \, , 
\label{v7}
\end{equation}
where ${\cal C}^{(\alpha )}$ is the universal function defined by (\ref{Cal-final},\ref{e8}, \ref{e9}). 
At this point all the numerical results in the previous subsection concerning ${\cal C}^{(\alpha )}$ apply.  

\bigskip 
\section{Schr\"odinger junction with harmonic potential}
\medskip 

In this section we consider a star graph $\Gamma$ with $M$ {\it infinite} edges 
and a harmonic potential $V(x) = \frac{1}{2} m \omega^2 x^2$ trapping the 
gas in the bulk (harmonic trap). The Schr\"odinger field $\psi_i(t,x)$ thus satisfies 
\begin{equation}
\left (i \partial_t + {1\over 2m} \partial_x^2 -\frac{1}{2} m \omega^2 x^2 \right )\psi_i (t,x) = 0
\, , \qquad x>0 \, . 
\label{hp1}
\end{equation} 
Since $V(x)$ defines a self-adjoint multiplication operator, the vertex boundary conditions 
controlling the self-adjointness of the total Hamiltonian $-\partial_x^2 + V(x)$ are still 
given by (\ref{bc1}). Accordingly, the critical $\S$ matrices are parametrized by (\ref{sd}). The 
field $\varphi_i(t,x)$, defined by (\ref{nl1}), satisfies 
\begin{eqnarray} 
\left (\partial_x \varphi_i\right )(t,0) &=& 0\, , \qquad 1\leq i \leq p\, , 
\label{hp2}\\
\varphi_i (t,0) &=& 0\, ,\qquad p< i \leq M\, , 
\label{hp3}
\end{eqnarray} 
for all $t$. The eigenfunctions of $-\partial_x^2 + V(x)$, obeying these boundary conditions, are 
\begin{equation} 
\phi_i(n,x)= 
\begin{cases} 
f_+(n,x)\, , & \qquad 1\leq i \leq p\, , \\ 
f_-(n,x)\, , & \qquad p< i \leq M\, ,
\end{cases} 
\label{hp4}
\end{equation}
where $f_\pm(n,x)$ are expressed in terms of the Hermite polynomials as follows\footnote{We set for simplicity 
$\omega =1$ and $m=1$.}
\begin{eqnarray} 
f_+(n,x) &=& \frac{1}{\pi^{1/4} \sqrt {2^{2n-1} (2n)!}}\, H_{2n}(x)\e^{-x^2/2}\, , \qquad \qquad n=0,1,...
\label{hp5}\\  
f_-(n,x) &=& \frac{1}{\pi^{1/4} \sqrt {2^{2n} (2n+1)!}}\, H_{2n+1}(x)\e^{-x^2/2}\, , \qquad \; n=0,1,...
\label{hp6}
\end{eqnarray} 
Inserting (\ref{hp5},\ref{hp6}) in ({\ref{v2}), one gets the two-point correlation function 
\begin{equation}
C^N_{ii}(x,y) = \sum_{n=0}^{2\N} {\overline \chi_i}(x,n) \chi_i (y,n) \, , 
\label{hp55}
\end{equation} 
with 
\begin{equation}
\chi_i (k,x)= 
\begin{cases} 
\frac{\Upsilon_i}{\pi^{1/4} \sqrt {2^{k-1}k !}}\, H_{k}(x)\e^{-x^2/2}\, , & \qquad k=1,3,...,2\N-1\, , \\ 
\\
\frac{\sqrt{1-\Upsilon_i^2}}{\pi^{1/4} \sqrt {2^{k-1}k !}}\, H_{k}(x)\e^{-x^2/2}\, , & \qquad k=0,2,...,2\N\, . \\ 
\end{cases}  
\label{hp66}
\end{equation} 

\begin{figure}[t]
\begin{center}
\includegraphics[scale=0.4]{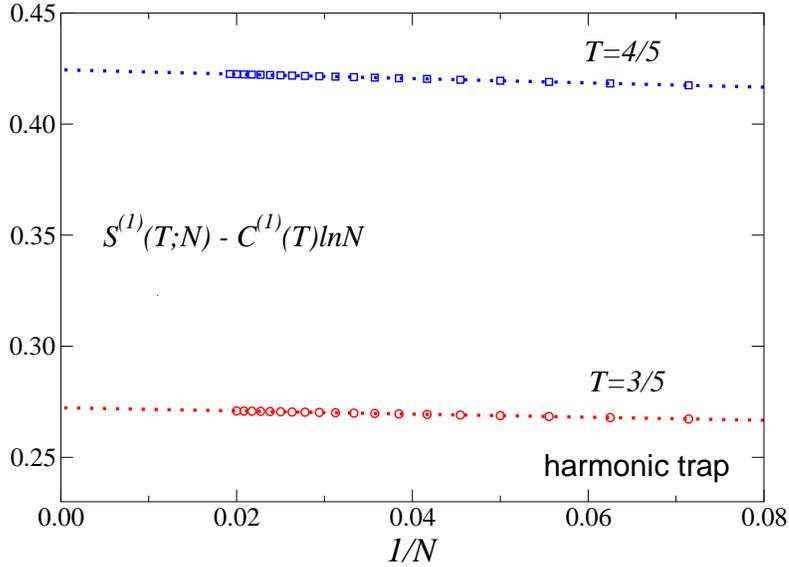}
\end{center}
\caption{ (Color online) The von Neumann edge entanglement entropy in
the two-wire junction with an external harmonic potential for $T=3/5$
and $T=4/5$.  We plot the subtracted quantity $S^{(1)}(T;N) - {\cal C}^{(1)}(T)\ln N$, 
where ${\cal C}^{(1)}(T)$ is given by
Eq.~(\ref{e8}).  The lines show fits of the data for $N\gtrsim 400$ to
a polynomial $b_0+b_1/N+b_2/N^2$.}
\label{junction}
\label{hbn2h}
\end{figure} 

At this point one can derive the relative $\A$-matrix and compute the entanglement entropy. 
The numerical data, displayed in fig. \ref{hbn2h}, confirm that in the harmonic case $S^{(\alpha)}_i(T_i;N)$ 
has precisely the behavior described by equation (\ref{v7}). 
An educated guess for the large-$N$ asymptotic behavior of the edge
entanglement entropy is that it is the same as that of the
hard-wall case, i.e.
\begin{equation}
S^{(\alpha)}(T;N) = {\cal C}^{(\alpha)}(T) \ln N + O(1),
\label{asbehb2h}
\end{equation}
where the functions ${\cal C}^{(\alpha)}$ are the same as those in
Eq.~(\ref{e8}).  Difference occurs at the level of the $O(1)$ term, as
in the case without defect~\cite{Calabrese:2011zz, cv-10, cv-10-2}.

In order to check it for nontrivial defects, we have computed the
entanglement entropy up to $N\approx 50$ for the two-edge problem, and
for two values of $T$, $T=3/5$ and $T=4/5$.  The results for the von
Neumann entropy are shown in Fig.~\ref{hbn2h}.  They are perfectly
consistent with the conjecture (\ref{asbehb2h}).  For example a fit of
the data for $T=4/5$ and even $N$ to
\begin{equation}
S^{(1)}(T;N) = a \ln N + b_0 + b_1/N + b_2/N^2 + b_3/N^2
\label{fitansh}
\end{equation}
give $a=0.118842$, $b_0=0.424421$, etc..., where $a$ should be
compared with ${\cal C}^{(1)}(T=4/5)=0.11884065...$.

\bigskip 
\section{Outlook and conclusions} 
\medskip

We have studied the entanglement entropies of one edge $i$ with respect to the rest of a junction with $M$ edges.
Our main result is that when working with a finite number of particles $N$ in edges of finite length $L$, 
the R\'enyi entanglement entropies can be derived analytically, obtaining
\be
S^{(\a)}= {\cal C}^{(\a)}(T)\ln N+ O(N^0)\,,
\ee
where $T$ is the total transmission probability from the edge $i$ to the rest of the graph
and the pre-factor is independent on the number of edges.
The analytical computation of ${\cal C}^{(\a)}$ is given by Eq. (\ref{Cal-final}) for integer $\a$
and its analytical continuation to non-intger $\a$ is given by the prefactor of 
Eisler and Peschel \cite{ep-10} obtained for the spatial entanglement of a line with a defect
reported in Eq. (\ref{e8}).
Clearly the value of the total transmission does depend on the number of edges and of the kind of junction. 
The same asymptotic behavior in $N$ also describe the entanglement entropies of systems in which on each arm 
there is a confining parabolic potential. 

We can turn the above asymptotic behavior in a more standard expression for the dependence of the entanglement 
entropies on the length of the subsystem (i.e. the edge in our case). Indeed, assuming a uniform density of 
particles we have $N\propto L$ and so
\be
S^{(\a)}= {\cal C}^{(\a)}(T)\ln L+ O(L^0)\,,
\label{Sasy}
\ee
that is expected to be valid also for lattice models and in particular for $M$ XX spin chains of length $L$
joined at a single common vertex.

Finally, we want to mention that the asymptotic result (\ref{Sasy}) has not yet been derived from conformal field theory.
Although it is clear that such derivation must be possible because the CFT encodes all the required 
ingredients, the practical calculation is very cumbersome, as the similar (but different) result of Ref. 
\cite{ss-08} shows. 
 
\bigskip

{\it Acknowledgements}. 

We are very grateful to Ingo Peschel for an intense correspondence about an independent and different 
way to obtain Eq. (\ref{Cal-final}) and for sharing the results of Ref. \cite{p-pc} before publication.
PC research was supported by ERC under the Starting Grant  n. 279391 EDEQS.
He also acknowledges the kind hospitality at the Institute Henri Poincar\'e, Paris.


\begin{thebibliography}{99} 

%%%%transport

\bibitem{kf-92} 
C.~L.~Kane and M.~P.~A. Fisher, Phys. Rev. Lett. {\bf 68}, 1220 (1992); 
Phys. Rev. {\bf B 46}, 15233 (1992). 

\bibitem{SS}
I.~Safi and H.~J.~Schulz, 
Phys.\ Rev.\  B {\bf 52}, R17040 (1995).

\bibitem{nfll-99} 
C. Nayak, M. P. A. Fisher, A.~W.~W.~Ludwig, and H.~H.~Lin, 
Phys.\ Rev.\  B {\bf 59}, 15694 (1999).

\bibitem{sdm-01} I. Safi, P. Devillard, and T. Martin, 
Phys. Rev. Lett. {\bf 86}, 4628 (2001). 



\bibitem{mw-02} J.E. Moore and X.-G. Wen, Phys. Rev. B {\bf 66}, 115305 (2002).

\bibitem{y-02} H. Yi, Phys. Rev. B {\bf 65}, 195101 (2002). 


%%%frameworks

\bibitem{lrs-02} 
S. Lal, S. Rao, and D. Sen, Phys. Rev. B {\bf 66}, 165327 (2002).

\bibitem{bac}
C. Bachas, J. de Boer, R. Dijkgraaf, and H. Ooguri,
JHEP 0206, 027 (2002).

\bibitem{rs-04} 
S. Rao and D. Sen, Phys. Rev. B {\bf 70}, 195115 (2004). 

\bibitem{coa-03} C. Chamon, M. Oshikawa, and I. Affleck,
Phys. Rev. Lett. {\bf 91}, 206403 (2003); 
M. Oshikawa,  C. Chamon, and I. Affleck,  J. Stat. Mech. P02008 (2006). 

\bibitem{emabms-05} T.~Enss, V.~Meden, S.~Andergassen, X.~Barnabe-Theriault,
W.~Metzner, K.~Schonhammer, Phys. Rev. B {\bf 71}, 155401 (2005);
X.~Barnabe-Theriault, A~ Sedeki, V.~Meden, K.~Schonhammer, 
Phys. Rev. Lett. {\bf 94}, 136405 (2005);
X.~Barnabe-Theriault, A.~Sedeki, V.~Meden, K.~Schonhammer, 
Phys. Rev. B {\bf 71}, 205327 (2005). 

\bibitem{ff-05}
D. Friedan, cond-mat/0505084; cond-mat/0505085. 

\bibitem{drs-06} S.~Das, S.~Rao, D.~Sen, Phys. Rev. B {\bf 74}  (2006) 045322.

%\cite{Bellazzini:2006jb}
\bibitem{Bellazzini:2006jb} 
B.~Bellazzini and M.~Mintchev, 
%``Quantum fields on star graphs,'' 
J.\ Phys.\ A  {\bf 39} (2006)  11101. 
%[arXiv:hep-th/0605036]. 
%%CITATION = JPAGB,A39,11101;%% 

%\cite{Bellazzini:2006kh}
\bibitem{Bellazzini:2006kh} 
B.~Bellazzini, M.~Mintchev and P.~Sorba,
%``Bosonization and scale invariance on quantum wires,'' 
J.\ Phys.\ A  {\bf 40} (2007) 2485. 
%[arXiv:hep-th/0611090].
%%CITATION = JPAGB,A40,2485;%%


\bibitem{hc-08}
C.-Y. Hou and C. Chamon, Phys. Rev. B {\bf 77} (2008) 155422. 

\bibitem{dr-08} S. Das and S. Rao, Phys. Rev. B {\bf 78}  (2008) 205421. 

%\cite{Bellazzini:2008fu}
\bibitem{Bellazzini:2008fu} 
B.~Bellazzini, P.~Calabrese and M.~Mintchev, 
Phys.\ Rev.\  B {\bf 79} (2009) 085122 
%[arXiv:0808.2719]. 
%%CITATION = PHRVA,B79,085122;%% 

%\cite{Bellazzini:2009nk}
\bibitem{Bellazzini:2009nk} 
B.~Bellazzini, M.~Mintchev and P.~Sorba, 
Phys.\ Rev.\  B {\bf 80}  (2009) 245441, 
%arXiv:0907.4221[hep-th].
%%CITATION = ARXIV:0907.4221;%% 


%\cite{Soori:2010ga}
\bibitem{Soori:2010ga}
A.~Soori and D.~Sen, 
Europhys.\ Lett.\  {\bf 93 } (2011) 57007. 


%%%%%%%Enta general

\bibitem{rev} L. Amico, R. Fazio, A. Osterloh, and V. Vedral, 
%Entanglement in many-body systems,
Rev. Mod. Phys. {\bf 80} (2008) 517;
J. Eisert, M. Cramer, and M. B. Plenio, 
%Area laws for the entanglement entropy - a review,
Rev. Mod. Phys. {\bf 82} (2010) 277;
%Entanglement entropy in extended systems,
P. Calabrese, J. Cardy, and B. Doyon Eds, J. Phys. A {\bf 42}  (2009) 500301.

\bibitem{sce}
J. Eisert and M. Cramer,
%Single-copy entanglement in critical spin chains,
Phys. Rev. A {\bf 72} (2005) 42112; 
I. Peschel and J. Zhao, %On single-copy entanglement,
J. Stat. Mech. (2005) P11002; %[quant-ph/0509002];
R. Orus, J.I. Latorre, J. Eisert, and M. Cramer,
%Half the entanglement in critical systems is distillable from a single specimen,
Phys. Rev. A {\bf 73} (2006) 060303. %[quant-ph/0509023].

\bibitem{cl-08}
P. Calabrese and A. Lefevre,
Phys. Rev. A 78, 032329 (2008).

\bibitem{c-lec}
J. Cardy,  %The ubiquitous 'c': from the Stefan-Boltzmann law to quantum information, 
J. Stat. Mech. (2010) P10004.

\bibitem{holzhey} C. Holzhey, F. Larsen, and F. Wilczek, 
%Geometric and renormalized entropy in conformal field-theory, 
Nucl. Phys. B {\bf 424}  (1994) 443.

\bibitem{vidalent} G. Vidal, J. I. Latorre, E. Rico, and A. Kitaev,
%Entanglement in quantum critical phenomena,
Phys. Rev. Lett. {\bf 90}  (2003) 227902.
J. I. Latorre, E. Rico, and G. Vidal,
%%Ground state entanglement in quantum spin chains,
Quant. Inf. Comp. {\bf 4} (2004) 048 .

\bibitem{cc-04}
P. Calabrese and J. Cardy, 
%Entanglement entropy and quantum field theory, 
J. Stat. Mech.  (2004) P06002.

\bibitem{cc-rev}
P. Calabrese and J. Cardy, 
%Entanglement entropy and conformal field theory,
J. Phys. A {\bf 42}  (2009) 504005. 



%%%Entanglement & Defects

\bibitem{l-04}
G. Levine, %Entanglement entropy in a boundary impurity model, 
Phys. Rev. Lett. {\bf 93}  (2004) 266402.

\bibitem{p-def} I. Peschel,
%Entanglement entropy with interface defects,
J. Phys. A: Math. Gen. {\bf 38}  (2005)  4327. 

\bibitem{def2}
J. Zhao, I. Peschel, and X. Wang,
%Critical entanglement of XXZ Heisenberg chains with defects,
Phys. Rev. B {\bf 73}, 024417 (2006);\\
J. Ren, S. Zhu, and X. Hao,
%Entanglement entropy in an antiferromagnetic Heisenberg spin chain with boundary impurities, 
J. Phys. B {\bf 42} (2009) 015504.



\bibitem{isl-09}
F. Igloi, Z. Szatmari, and Y.-C. Lin,
%Entanglement entropy with localized and extended interface defects,
Phys. Rev. B {\bf 80}  (2009) 024405.

\bibitem{ss-08}
K. Sakai and  Y. Satoh, %Entanglement through conformal interfaces,
JHEP 0812: 001 (2008).

\bibitem{scla-07}
E. S. Sorensen, M.-S. Chang, N. Laflorencie, and I. Affleck,
%Impurity Entanglement Entropy and the Kondo Screening Cloud,
J. Stat. Mech. (2007) L01001; J. Stat. Mech. (2007) P08003;
E. S. Sorensen, N. Laflorencie, and I. Affleck,
J. Phys. A {\bf 42}, 504009 (2009). 

\bibitem{ep-10} 
V. Eisler and I. Peschel, 
Ann. Phys. (Berlin) {\bf 522} (2010) 679.

\bibitem{p-11}
I. Peschel,  1109.0159.

\bibitem{eg-10}
V. Eisler and S. Garmon. Phys. Rev. B 82, 174202 (2010).

%%%%%entanglement

%\cite{Calabrese:2011zz}
\bibitem{Calabrese:2011zz} 
P.~Calabrese, M.~Mintchev, and E.~Vicari, 
%``Entanglement Entropy of One-Dimensional Gases,'' 
Phys.\ Rev.\ Lett.\  {\bf 107 } (2011)  020601. 

%\cite{Calabrese:2011vh} 
\bibitem{Calabrese:2011vh} 
P.~Calabrese, M.~Mintchev, and E.~Vicari, 
%``The Entanglement entropy of 1D systems in continuous and homogenous space,'' 
 J. Stat. Mech. (2011) P09028. 
 
\bibitem{cmv-11}
P.~Calabrese, M.~Mintchev, and E.~Vicari, 
%Exact relations between particle fluctuations and entanglement in Fermi gases, 
1111.4836.


%%%%q-graphs


\bibitem{ks-00}
V.~Kostrykin and R.~Schrader, 
Fortschr. Phys. {\bf 48} (2000) 703. 

\bibitem{h-00}
M.~Harmer, 
J.\ Phys.\ A {\bf 33} (2000) 9015. 


%%%%%%%%%%%

\bibitem{jk-04}
B-Q Jin and V.E. Korepin, J. Stat. Phys. {\bf 116} (2004) 79. 

\bibitem{ep-rev}
I. Peschel and V. Eisler, J. Phys. A {\bf 42}  (2009) 504003.

\bibitem{Klich-06} I. Klich, %Lower entropy bounds and particle number fluctuations in a Fermi sea,
 J. Phys. A {\bf 39}, L85 (2006).
 
\bibitem{p-pc}
I. Peschel and V. Eisler, 1201.4104. 
 
\bibitem{corrlatt}
N Laflorencie, E S Sorensen, M-S Chang, and I Affleck,
%Boundary effects in the critical scaling of entanglement entropy in 1D systems,
Phys. Rev. Lett. {\bf 96}, 100603 (2006);
 P. Calabrese, M. Campostrini, F. Essler, and B. Nienhuis,
%Parity effects in the scaling of block entanglement in gapless spin chains, 
Phys. Rev. Lett. {\bf 104}, 095701 (2010);
P. Calabrese and F. H. L. Essler,
%Universal corrections to scaling for block entanglement in spin-1/2 XX chains
J. Stat. Mech. (2010) P08029; 
M. Fagotti and P. Calabrese, 
%{Universal parity effects in the entanglement entropy of XX chains with open boundary conditions},
J. Stat. Mech. P01017 (2011).

\bibitem{cv-10}
M. Campostrini and E. Vicari, J. Stat. Mech. (2010) P08020;
J. Stat. Mech. (2011) E04001;
Phys. Rev. A {\bf 81}  (2010) 063614.

\bibitem{cv-10-2}
M. Campostrini and E. Vicari,
Phys. Rev. A {\bf 82}  (2010) 063636. 




%%%% References not yet quoted in the text %%%%%%%%%%%%%%



%\bibitem{mps} J.I. Cirac and F. Verstraete,  J. Phys. A {\bf 42} (2009) 504004.


%\bibitem{vc-10} F. Verstraete and J.I. Cirac, Phys. Rev. Lett. {\bf 104}  (2010) 190405.



%\bibitem{bk11}
%F.N.C~Paraan {\em et al.},
%J. Molina-Vilaplana, V. E. Korepin, and S. Bose, 
%1105.1211.

%\bibitem{QH}
%M. Haque, O. Zozulya, and K. Schoutens, 
%Phys. Rev. Lett. {\bf 98}, 060401 (2007);
%H. Li and F. D. M. Haldane, Phys. Rev. Lett. {\bf 101}, 010504 (2008).

%\bibitem{mzs-09}
%M. Haque, O. Zozulya, K. Schoutens, J. Phys. A {\bf 42}, 504012 (2009).







%\bibitem{fh} The Fisher-Hartwig conjecture [M. E. Fisher and
%R. E. Hartwig, Adv. Chem. Phys. {\bf 15}, 333 (1968)] has been
%rigorously proven for the case at hand, see e.g. E. L. Basor and
%K. E. Morrison, Lin. Alg.  Appl. {\bf 202}, 129 (1994).



 
\end{thebibliography}
\end{document}